\documentclass[fleqn,twoside,aps]{article}
\usepackage{espcrc2}
\usepackage{graphicx}
\usepackage{epsfig}

\newcommand{\AmS}{{\protect\the\textfont2
  A\kern-.1667em\lower.5ex\hbox{M}\kern-.125emS}}
\title{Kaon and Pion Electromagnetic Form Factor Ratios in the Light-Front}

\author{O. A. T. Dias\address[IFT]{Instituto de F\'\i sica Te\'orica,
Universidade Estadual Paulista, IFT-UNESP, Rua Dr. Bento Teobaldo Ferraz,
271 Bl. II  Barra Funda, 01140-070, S\~ao Paulo, SP},
V. S. Filho\address[LFTC]
{Laborat\'orio de F\'\i sica Te\'orica e Computa\c c\~ao
Cient\'\i fica, LFTC, Universidade Cruzeiro do Sul \\ 01506-000,
S\~ao Paulo, SP., Brazil}  and
J.~P.~B.~C.~de~Melo\addressmark[IFT] \addressmark[LFTC]}

\begin{document}

\begin{abstract}
We have applied the light-front formalism to calculate the
electromagnetic form factors for the pion and the kaon from two
models at low and high energies in order to explore the
differences between such models. We have also compared the results
for the ratio $F_{K}(Q^2)/F_{\pi}(Q^2)$ with the experimental data
up to 10 [GeV/c]$^2$ and we have observed that the theoretical
results are in good concordance for low energies, but they are
very different at higher energy scales. \vspace{1pc}
\end{abstract}
\maketitle

\section{Introduction}
An interesting laboratory for studying the structure of elementary 
particles is the electromagnetic form factor $F_{\pi}(q^2)$ of the
pion, as there are a lot of experimental data for this
observable~\cite{Data}. In the case of the $K^+$ kaon, there is 
much less knowledge about its electromagnetic form factor
$F_K(q^2)$, but it is possible to find some experimental results for
$q^2>1~{\rm [GeV/c]}^2$~\cite{new}. Further, the recent
high-statistics Brookhaven experiment E865~\cite{Appel} has
provided us with more information on $F_K$. The amplitude for the
decay $K^+\rightarrow\pi^+e^+e^-$ was measured for $q^2$ up to
0.125 [GeV/c]$^2$ (the maximum value for the kinematics of this
kaon decay), that allows to obtain indirectly the value of
$F_K(q^2)$~\cite{Lowe}.

The main objective of the light-front models used in this work is
to describe consistently  hadronic bound state systems for high
and low $Q^2$ regimes. In light-front models, the bound state wave
function is defined on the hypersuperface $x^+=x^0+x^3=0$ and it
is covariant under kinematical boosts, due to the stability of
Fock-state decomposition under such boosts~\cite{Perry90}.

As described in Ref.~\cite{Pacheco,Pacheco99}, problems related to
the rotational symmetry breaking make the results of
electromagnetic form-factor calculations in the light-front
formalism dependent on which component of the electromagnetic
current is used to obtain the form-factors.

At low momentum transfers, the non-perturbative regime of QCD is
more important when compared with high momentum transfers
dominated by the perturbative regime of QCD. Perturbative QCD
works well after $1.0$~[GeV/c]$^2$ and dominates above
$5.0$~[GeV/c]$^2$.

The studies on light-flavor vector and scalar mesons are important
because they indicate a direction to understand why QCD works in
the non-perturbative regime and, also, why the pseudoscalar mesons
are the observed light hadrons related with  chiral symmetry
breaking.

As known for spin-1 particles~\cite{Pacheco}, the plus
component~("$J^{+}$") of the electromagnetic current is not free
from pair term contributions in the Breit frame ($q^+=0$), so the
rotational symmetry is broken if they are dropped out. Thus, the
matrix elements of the electromagnetic current in the light-front
formalism have the valence contribution to the electromagnetic
current but also other contributions should be considered. That contribution
corresponds to pair terms added to the matrix elements of the
electromagnetic current~\cite{Pacheco}.

In the present work, one type of the vertex function is used in
order to calculate the pion electromagnetic form-factor for the 
$\pi-q\bar{q}$ vertex. The light-front models for the pion and the 
kaon which were presented at previous works~\cite{Pacheco} are 
applied to high momentum transfers. 

In section II, we present the formalism for the decay constants 
and electromagnetic form factor of the pseudoscalar mesons for a 
nonsymmetric vertex model. In our calculations, we also use a  
symmetric vertex model of the pion and kaon given in this section. 
In the last section, we show our numerical results for the ratio
between the kaon and pion electromagnetic form factors.  We also
discuss the main relevant points of the results.
\section{Electromagnetic Form Factor for Pseusoscalar Mesons}
\vskip .25 cm

The electromagnetic form factor for pseudoscalar particles can be
obtained from a covariant expression as:
\begin{equation}
<p|J^{\mu}|p^{\prime}> = (p^{\prime} + p)^{\mu} F_{PS}(q^2),
\end{equation}
where $J^\mu$ is the electromagnetic current and $F_{PS}(q^2)$ is
the pseudoscalar electromagnetic form factor.

The pseudoscalar pion decay constant is given by
\begin{eqnarray}
i p^\mu f_{PS} = \frac{m}{f_{PS}}
N_c \int{\frac{dk^4}{(2 \pi^4)}  Tr[ {\cal{O}}  ]
\Lambda_M(k,p)},
\end{eqnarray}
where
\begin{eqnarray}
{\cal{O}}
=\gamma^\mu \gamma^5 S(k)\gamma^5 S(k - p). \nonumber
\end{eqnarray}

\subsection{Pion Form Factor}
\vskip .25 cm The pion electromagnetic current can be written as:
\begin{equation}
J^{\mu}=e(p^{\mu}+p^{\prime\mu})F_{\pi}(q^2). \label{current}
\end{equation}
In the equation above, in general, it is possible to extract the
form factor using either the plus or minus components of the
electromagnetic current, $J^{+}$ and $J^{-}$, respectively. In a
reference frame, where the plus component of the moment transfer
$q^+=q^0+q^3$ is nonzero, the electromagnetic form factor has two
contributions:
\begin{eqnarray}
F_{\pi}(q^2)=F^{(I)}_{\pi}(q^2)+F^{(II)}_{\pi}(q^2),
\label{ffactor}
\end{eqnarray}
where $F^{(I)}_{\pi}(q^2)$ and $F^{(II)}_{\pi}(q^2)$ are the
valence and the non-valence terms, respectively.

\begin{table}[h]
\caption{Parameters for the nonsymmetric vertex model (NSM) and
the light-front covariant model (LFCM). The scale parameters $\lambda_\pi$ and
$\lambda_K$ are fitted to the corresponding decay constants.}
\begin{tabular}{|ll|}
\hline
NSM & \\
\hline
$m_{u}$  & $0.220$~GeV \\

$m_{\bar{s}}$ &  $0.419$ GeV \\

$m_R$  &  $ 0.946$ GeV\\

$f_{\pi}$ & $0.101$ GeV\\

$ m_{\pi} $ &  $0.140$ GeV \\

$m_{K^+}$  &  $0.494$ GeV\\

 $\langle r_{\pi^+} \rangle$  & $0.67fm$\\
\hline
LFCM $(n=2)$ & \\
\hline
$m_u$  & $0.220$ GeV \\

 $m_{\pi^+}$ & $0.140$ GeV \\

$\lambda_{\pi}$ & $0.542$ GeV \\

 $\langle r_{\pi^+} \rangle $  & $ 0.576$ fm \\

$f_{\pi^+}$ & $0.0924$ GeV \\

 $m_{\bar{s}}$  &  $0.344$ GeV \\

$m_{K^+}$ & $0.454$ GeV \\

$\lambda_{K}$ & $0.621$ GeV \\

$\langle r_{K^+} \rangle $ &  $0.513$ fm \\

 $f_{K^+}$ & $0.113$ GeV\\
\hline
\end{tabular}
 \end{table}

From the analytical integration of the $k^-$ loop momentum in
Eq.~(\ref{ffactor}), one finds two possible intervals of $k^+$
which give nonzero contribution: (i) $0 < k^+ < p^+$ and (ii)
$p^{+} < k^+ < p^{\prime+}$. The interval (i) corresponds to the
valence contribution to the electromagnetic current and the second
interval (ii) is the non-valence contribution. The sum of both
contributions gives the covariant form factor, which depends only
on $q^2$ and not on the particular value of $q^+$.

The electromagnetic form factor $F^{{(NSY)}}_{\pi}(q^2)$ for the
pion can be expressed with the nonsymmetric vertex model~(see the
reference~\cite{Pacheco99} for details) as:
\begin{eqnarray}
F_{\pi}(q^2)=\imath e
\frac{m^2 N^2}{p^+ f^2_\pi} N_c \int \frac{
d^{2} k_{\perp} d k^{+}}{2 (2 \pi)^4} I(k_{\perp},k^{+}),
\end{eqnarray}
where
{\small{
\begin{eqnarray}
I=\frac{M\Theta}{k^+(p^+-k^+)^2(p^{\prime +}-k^+)^2}    \nonumber
\end{eqnarray}}}
with $M$ given by:
\begin{eqnarray}
 -4\bar{k}^-k^-(k^+-p^+)^2+4(k^2_\perp+m^2)(k^+-2p^+)+k^+q^2,
 \nonumber
\end{eqnarray}
\begin{eqnarray}
&&\Theta = \frac{\theta(k^+) \theta(p^+ -
k^+)}{H(k_{\perp},k^{+})},~\bar{k}^-=\frac{f_1-\imath
\epsilon}{k^+}, \nonumber
\end{eqnarray}
and {\small
\begin{eqnarray}
H=(p^- - \bar{k}^- - \frac{f_2 -\imath \epsilon }{p^+ -
k^+})(p^{\prime-}-\bar{k}^--\frac{f_3-\imath
\epsilon}{p^{\prime+}-k^+})\times \nonumber \\
(p^{-}-\bar{k}^--\frac{f_4-\imath\epsilon}{p^{+}-k^+})
(p^{\prime-}-\bar{k}^--\frac{f_5-\imath\epsilon}{p^{\prime+}-k^+}).
\nonumber
\end{eqnarray}}
The function $f_i$ for $1\leq i \leq 5$  is defined as
$r^2_\perp+a^2$, with
$(r,a)=(k,m);(p-k,m);(p^\prime-k,m_R);(p-k,m);(p^\prime-k,m_R)$,
respectively to the sequence 1 to 5. The constituent quark mass is
$m$ and the regulator mass is $m_R$ given in Table I.

\subsection{Kaon Form Factor}
\vskip .25cm
The expression for the kaon electromagnetic form factor is given by:
\begin{equation}
<p|J^{\mu}|p^{\prime}> = e(p^{\prime\mu} + p^{\mu}) F_K(q^2),
\end{equation}
where $J^\mu$ is the electromagnetic current and $F_K(q^2)$ is the kaon form factor.

The electromagnetic form factor of the kaon
\begin{eqnarray}
F_K(q^2)=F_{q}(q^2)+F_{\bar q}(q^2) \ ,
\end{eqnarray}
receives contribution form the quark and antiquark currents. With
the nonsymmetric vertex~\cite{Fabiano} in the light-front
coordinates the contribution of the quark current to the form
factor can be expressed as:
\begin{eqnarray}
F_{q}(q^2)  =  2 \imath e_{q} \frac{N^2 g^2 N_c}{2 p^{+}} \int
\frac{d^{2} k_{\perp} d k^{+} d k^{-} }{2 \ (2 \pi)^4} \ \ I
\end{eqnarray}
where $I=I_1(k_{\perp},k^{+},k^{-})\times I_2(k_{\perp},k^{+},k^{-})$ and
the functions $I_1$ and $I_2$ are given by:
\begin{eqnarray}
I_1=\frac{A(k_{\perp},k^{+},k^{-})}
{k^+(p^+-k^+)^2(p^{^{\prime}+}-k^+)^2(k^--\frac{f_6-\imath\epsilon}{k^+})}
\nonumber
\end{eqnarray}
where $f_6=k^2_\perp+m_{\bar q}^2$ and
\begin{eqnarray}
I_2=\frac{-4k^+m^{2}_{q}+8(k^+-p^+)m_{q}m_{\bar{q}}}
{B(k_{\perp},k^{+},k^{-})},  \nonumber \label{ffactor+}
\end{eqnarray}
with
\begin{eqnarray}
 A=-4(k^{-}k^{+2}-k^{+}k^2_\perp-2k^{-}k^+p^+  \nonumber \\
 +2k_\perp p^+-\frac{k^{+}q}{4}+k^-p^{+}). \nonumber
\end{eqnarray}
and
{\small
\begin{eqnarray}
B=(p^--k^{-}-\frac{f_2-\imath\epsilon}{p^+-k^+})
(p^{\prime-}-k^{-}-\frac{f_3-\imath\epsilon}{p^{\prime+}-k^+}) \times\nonumber \\
(p^--k^{-}-\frac{f_4-\imath\epsilon}{p^+-k^+})
(p^{\prime-}-k^{-}-\frac{f_5-\imath\epsilon}{p^{\prime+}-k^+}) .
\nonumber
\end{eqnarray}}
The functions $f_2$ to $f_5$ in the equation above are defined as
previously replacing $m$ by $m_q$.

\begin{figure}[h]
\vspace{-0.5cm}
\includegraphics[angle=-90, scale=.345]{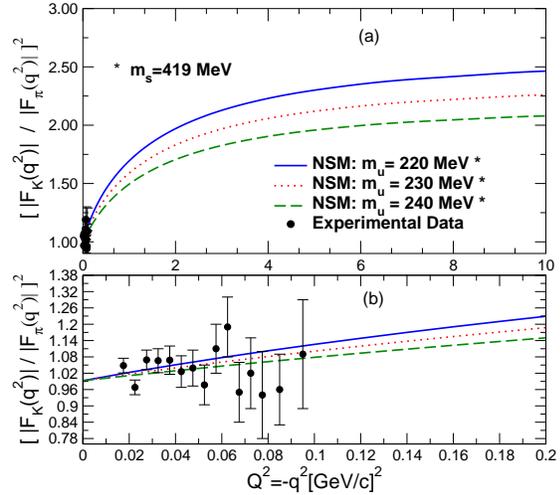}
\vspace{-1.2cm} \caption{(a) Ratio between the kaon and pion form
factors for the nonsymmetric model for Q$^2$ up to
10~$[GeV/c]^2$; (b) calculations for Q$^2$ up to 0.2~$[GeV/c]^2$
compared to the experimental data~\cite{Data}.} \label{lcp1}
\vspace{-.5cm}
\end{figure}

\begin{figure}[h]
\includegraphics[angle=-90, scale=.345]{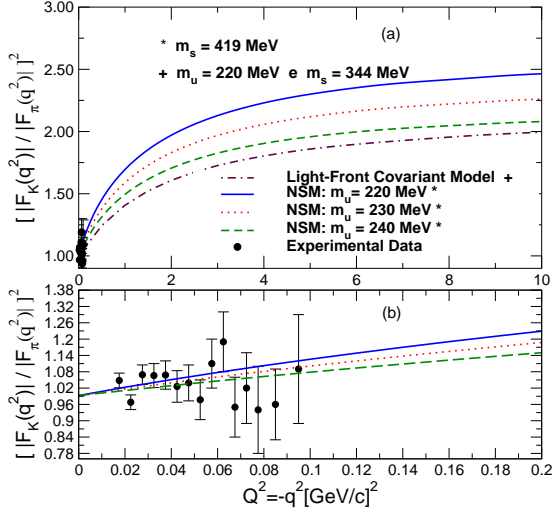}
\vspace{-1.2cm} \caption{(a) Ratio between the kaon and pion form
factors for the Light-Front Covariant Model up to Q$^2$=
10~$[GeV/c]^2$; (b) calculations for Q$^2$ up to 0.2~$[GeV/c]^2$
compared to the experimental data~\cite{Data}.} \label{lcp2}
\end{figure}

\subsection{Light-Front Covariant Model:
The Bethe Salpeter Amplitude}
\vskip .25cm
The Bethe-Salpeter amplitude is calculated with
the light-front covariant model~\cite{POS2008}:
{\small{
\begin{eqnarray}
\Lambda_M \left( k, p \right) = \frac{( k_1^2 - m_1^2) \Gamma_M
(k_2^2 -
m_2^2)}{(k_1^2-\lambda_M^2+i\epsilon)^n(k_2^2-\lambda_M^2+i\epsilon)^n},
\end{eqnarray}}}
where $k_1=k$ and $k_2=p-k$. $\lambda_M$ is the scale associated
with the meson light-front valence wave function, $n$ is the power
of the regulator and $m_1$ and $m_2$ are the quark masses within
the meson bound state.

\section{Results and Conclusions}
The parameters of our models, NSM and LFCM, are given in the Table
I. In figure \ref{lcp1}, we show the numerical results of the
ratio between the kaon and pion electromagnetic form factors
obtained with NSM compared to the experimental data. In figure
\ref{lcp1}(a) (upper frame), we show our results for Q$^2$ up to
10~$[GeV/c]^2$. In order to show in detail the comparison with
available experimental data, it is presented the ratio for Q$^2$
up to 0.2~$[GeV/c]^2$, in figure \ref{lcp2}(b)(lower frame).

We show in figure \ref{lcp2}(a) calculations for the ratio using
the meson Bethe-Salpeter amplitude defined with the vertex given
by LFCM. We compare that to the calculations with NSM up to 10~$[GeV/c]^2$. 
From the figure, it is clear the sensitivity of the ratio to the 
different models. From figure \ref{lcp2}(b), 
we realize that both models can describe the
experimental data up to Q$^2$= 0.2~$[GeV/c]^2$. We conclude by
observing both frames in figure \ref{lcp2} that the two models are
in good agreement with the experimental data for low momentum and
show significant dependence on the quark mass for high momentum 
transfers, which can be clearly seen for the NSM. 
This is possibly correlated to modification of the pion and kaon decay constant
when changing the constituent quark mass, which appears at high
momentum transfers as the wave functions of the pion and kaon at
short distances are essentially defined by $f_\pi$ and $f_K$,
respectively.

We thank to FAPESP for partial financial support.

\end{document}